\documentclass{emulateapj}

\usepackage{apjfonts}
\usepackage[]{natbib}

\long\def\Ignore#1{\relax}

\newcommand{\kpc}{{\rm kpc}}
\newcommand{\Rc}{R_{\rm c}}

\newcommand{\Vc}{V_{\rm c}}

\newcommand{\Ms}{M_\sun}

\shorttitle{The Milky Way as a Pure-Disk Galaxy}

\shortauthors{Shen et al.}

\begin{document}
\title{Our Milky Way as a Pure-Disk Galaxy -- A Challenge for Galaxy Formation}

\author{Juntai Shen\altaffilmark{1,2}}
\author{R. Michael Rich\altaffilmark{3}}
\author{John Kormendy\altaffilmark{2}}
\author{Christian D. Howard\altaffilmark{3,4}}
\author{Roberto De Propris\altaffilmark{5}}
\author{Andrea Kunder\altaffilmark{5}}

\altaffiltext{1}{Key Laboratory for Research in Galaxies and Cosmology,
  Shanghai Astronomical Observatory, Chinese Academy of Sciences, 80 Nandan
  Road, Shanghai 200030, China}
\altaffiltext{2}{Department of Astronomy, The University of Texas at Austin, 1 University Station, C1400, Austin, Texas 78712, USA}
\altaffiltext{3}{Department of Physics and Astronomy, University of California
  at Los Angeles, Los Angeles, California, USA}
\altaffiltext{4}{NASA Ames Research Center, MS 211-3 Moffett Field, CA 94035, USA}
\altaffiltext{5}{Cerro Tololo Inter-American Observatory, Casilla 603, La Serena, Chile}

\slugcomment{Accepted by ApJL}

\begin{abstract}
Bulges are commonly believed to form in the dynamical violence of galaxy
collisions and mergers. Here we model the stellar kinematics of the Bulge
Radial Velocity Assay ({\sl BRAVA}), and find no sign that the Milky Way
contains a classical bulge formed by scrambling pre-existing disks of stars in
major mergers. Rather, the bulge appears to be a bar, seen somewhat end-on, as
hinted from its asymmetric boxy shape.  We construct a simple but realistic
$N$-body model of the Galaxy that self-consistently develops a bar. The bar
immediately buckles and thickens in the vertical direction. As seen from the
Sun, the result resembles the boxy bulge of our Galaxy. The model fits the
{\sl BRAVA} stellar kinematic data covering the whole bulge strikingly well
with no need for a merger-made classical bulge. The bar in our best fit model
has a half-length of $\sim 4\;\kpc$ and extends $20^\circ$ from the
Sun-Galactic Center line. We use the new kinematic constraints to show that
any classical bulge contribution cannot be larger than $\sim$ 8\% of the disk
mass. Thus the Galactic bulge is a part of the disk and not a separate
component made in a prior merger. Giant, pure-disk galaxies like our own
present a major challenge to the standard picture in which galaxy formation is
dominated by hierarchical clustering and galaxy mergers.
\end{abstract}


\section{Introduction}
\label{sec:intro}

Astronomers commonly make the Copernican assumption that our Milky Way is in
no way unusual; then they can exploit the fact that we live in it to study
galaxy formation in special detail. Past assumptions about our Galactic bulge
grew out of our developing understanding of galaxy formation.  It is well
known that spiral galaxies consist of three main components, an invisible dark
matter halo, an embedded, flat disk, and a central bulge.  The bulge of our
Galaxy is $>$99\% made of stars that are at least 5 Gyr old
\citep{cla_etal_08} with a wide range of metal abundances
\citep{mcw_ric_94,ful_etal_06,zoc_etal_08}.  In this respect
and many others, big bulges are similar to (diskless) elliptical galaxies.
The formation of ellipticals is well understood.  Hierarchical gravitational
clustering of initial fluctuations in the cosmological density results in
galaxy collisions and mergers that scramble flat disks into rounder
ellipticals \citep{toomre_77_merger,whi_ree_79,ste_nav_02,nak_nom_03}.  Significant
energy has been invested in developing this very successful theory of galaxy
formation, and it was natural to think that our Galactic bulge is a product of
it.  There is little danger that the picture is fundamentally wrong
\citep{binney_04_sydney}.

But it is incomplete.  The theme of this paper is that our Galactic bulge is
indeed normal but that it is prototypical of different formation processes
than are usually assumed.  A complementary suite of evolution processes shapes
isolated galaxies.  They evolve by rearranging energy and angular momentum;
this grows central components that masquerade as classical bulges but that
formed directly out of disks without any collisions
\citep{kormen_93,kor_ken_04}.  To distinguish them from merger remnants, we
call them ``pseudobulges''.  They come in two varieties.  Some are flattened;
they are grown out of disk gas transported inward by nonaxisymmetries such as
bars.  Another variety is recognized only in edge-on galaxies.  When bars form
out of disks, they buckle vertically and heat themselves into thickened
structures that look box-shaped when seen edge-on
\citep{com_san_81,com_etal_90,rah_etal_91}.  Our Galaxy contains such a
box-shaped \citep{mai_etal_78,wei_etal_94,dwe_etal_95} pseudobulge
\citep{kor_ken_04}.

The identification of this boxy structure as an edge-on bar is particularly
compelling because infrared imagery shows a parallelogram-shaped
distortion \citep{mai_etal_78,wei_etal_94,dwe_etal_95} that is naturally
explained as a perspective effect: the near end of the bar is closer to us
than the far end. So its vertical extent is taller on the near side than on
the far side \citep{bli_spe_91}. \citet{zhao_96} developed the first rapidly
rotating bar model that fitted this distortion.  Zhao's model was based on the
\citet{schwar_79} orbit superposition technique, so it was self-consistent and
in steady state, but it did not evolve into that state from plausible initial
conditions.  Also, little stellar kinematic data were available to constrain
Zhao's steady-state model and early $N$-body models
\citep{fux_97,fux_99,sev_etal_99}, and subsequent radial velocity data
from a survey of planetary nebulae, although compared with a range of
dynamical models \citep{bea_etal_00}, led to
only limited conclusions because of the small numbers and uncertain population
membership of the planetary nebulae.

In this paper, we simulate numerically the self-consistent formation of a bar
that buckles naturally into a thickened state, and we scale that model to fit
new kinematic data on bulge rotation and random velocities.  The radial
velocity observations are provided by the Bulge Radial Velocity Assay ({\sl BRAVA};
\citealt{ric_etal_07,how_etal_08}).  This is a spectroscopic survey of the
stellar radial velocities of M-type giant stars whose population membership in
the bulge is well established.  These giants
provide most of the 2 $\mu$m radiation whose box-shaped light
distribution motivates bar models.  {\sl BRAVA} emphasizes measurements in
two strips at latitude $b=-4^\circ$ and $b=-8^\circ$ and at longitude
$-10^\circ$ $< l <$ $+10^\circ$.  A strip along the minor axis ($l
\equiv 0^\circ$) has also been observed.  We use nearly 5,000 stellar
radial velocities in this report.  A preliminary analysis of data
found strong cylindrical rotation \citep{how_etal_09} consistent with an
edge-on, bar-like pseudobulge, although a precise fit of a bar model
to the data was not available. This success leads us here to construct
a full evolutionary $N$-body model that we can fit to the radial
velocity data.


\begin{figure}
\centerline{
\includegraphics[angle=0.,width=\hsize]{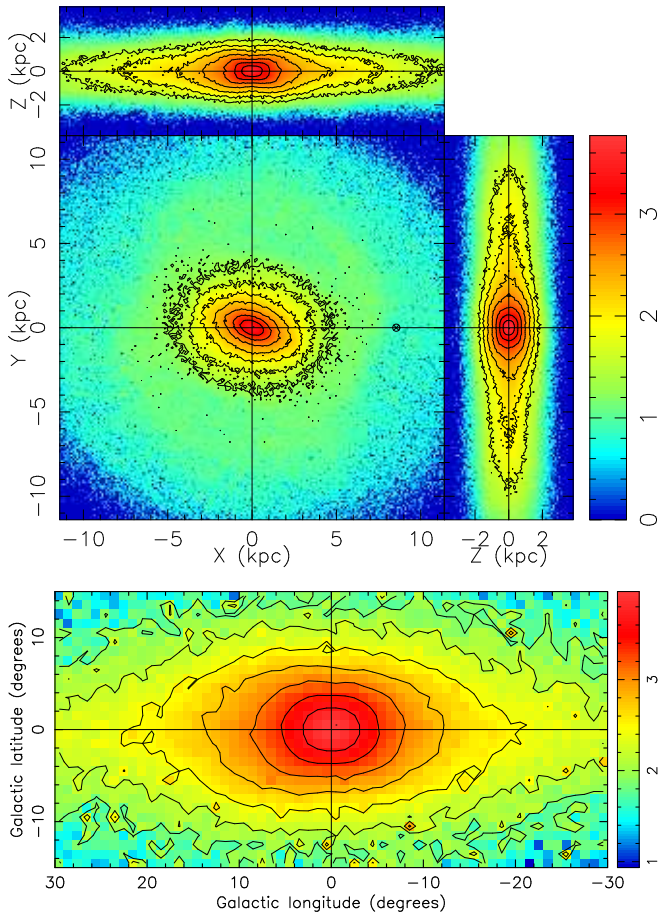}
}
\caption{Upper three panels: Face-on and side-on views of the surface density
  of our best-fitting model as seen from far away.  The Sun's position 8.5~kpc
  from the Galactic center is marked along the $+x$ axis.  The Galaxy rotates
  clockwise as seen in the face-on projection. Bottom panel: Model surface
  brightness map in Galactic coordinates as seen from the Sun's location.  Our
  perspective makes the box-shaped, edge-on bar look taller on its nearer
  side.  The Galactic boxy bulge is observed to be similarly distorted.}
\label{fig:cont}
\end{figure}

\section{Model Setup}

We use a cylindrical particle-mesh code \citep{sel_val_97, she_sel_04} to
build fully self-consistent $N$-body galaxies.  It is well suited to study the
evolution of disk galaxies: we model the disk with at least 1 million
particles to provide high particle resolution near the center where the
density is high.  We try to construct the simplest self-consistent $N$-body
models that fit the {\sl BRAVA} data, avoiding contrived models with too many free
parameters.  Initially, they contained only an unbarred disk and a dark
halo.  The profile of the Galactic halo is poorly constrained observationally;
we adopt a rigid pseudo-isothermal halo potential \hbox{$\Phi=\frac{1}{2}
  \Vc^2 \ln (1+\frac{r^2}{\Rc^2})$}. Here $\Vc$ $\sim$ 250 km s$^{-1}$ is the
asymptotic circular-orbit rotation velocity at infinity, and $\Rc=15$ kpc is
the core radius inside which the potential is effectively constant. This halo
gives a nearly flat rotation curve of $\sim$ 220 km~s$^{-1}$ between 5 to 20
kpc.  A simple halo form allows us to run many simulations quickly; this is
important for a parameter search such as the present one.  A rigid halo also
omits dynamical friction on the bar, but the central density of the cored halo
we adopt is low enough so that friction will be very mild
\citep[e.g.][]{deb_sel_00,ath_mis_02}. More importantly, we are mainly
interested in the bulge, which is embedded well interior to $\Rc$.  So the
exact profile of the dark halo at large radii is not critical.
We will explore more sophisticated halos in a future study.

\begin{figure}[!th]
\centerline{
\includegraphics[angle=0.,width=\hsize]{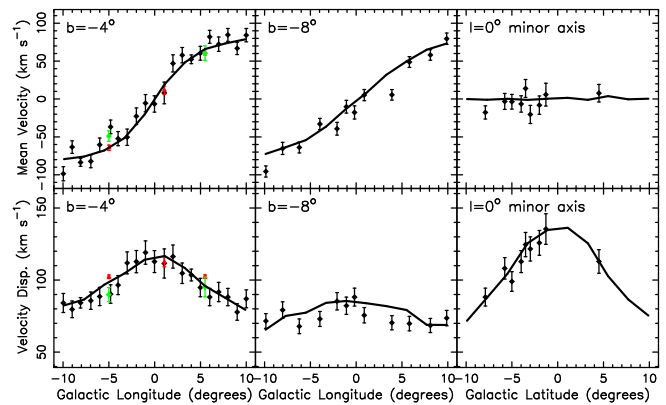}
}
\caption{(top): Mean velocity and velocity dispersion profiles of the
best-fitting model (black lines) compared to all available kinematic
observations.  The left two panels are for the Galactic latitude
$b=-4^\circ$ strip; the middle two panels are for the $b=-8^\circ$
strip; and the right two panels are for the $l=0^\circ$ minor
axis. The black diamonds and their error bars are the {\sl BRAVA} data; the
green diamonds are for M-type giant stars \citep{ran_etal_09}, and the
red triangles are the data on red clump giant stars \citep{ran_etal_09}.
This is the first time that a single dynamical model has been compared
with data of such quality.  The agreement is striking.}
\label{fig:fits}
\end{figure}

\begin{figure}[!h]
\centerline{
\includegraphics[angle=0.,width=.9\hsize]{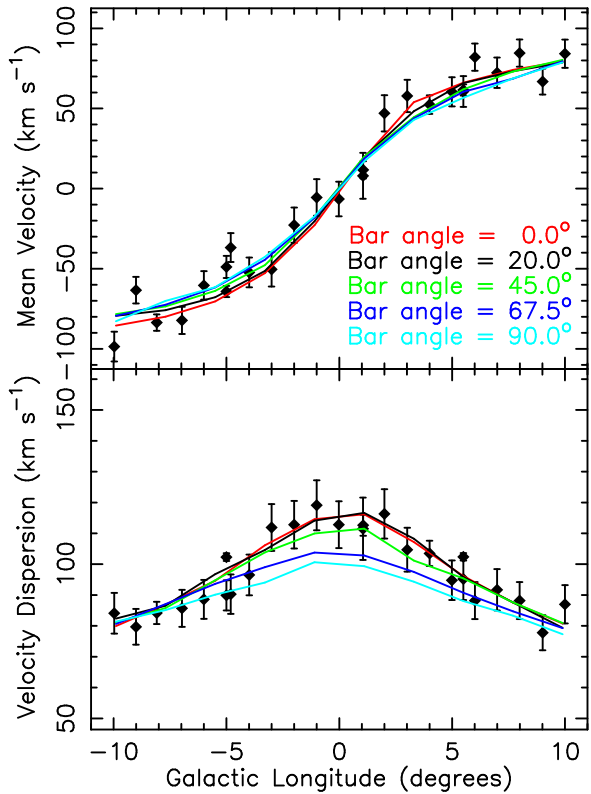}
}
\caption{Model results compared to the data from the $b=-4^\circ$
major-axis strip as we vary the bar angle relative to the line that
connects the Sun to the Galactic center. The red, black, green, blue,
and cyan curves are the model results for bar angles of $0^\circ$,
$20^\circ$ (the adopted best-fit value), $45^\circ$, $67.5^\circ$, and
$90^\circ$, respectively.  All data points are plotted as black
diamonds regardless of source. Clearly bar angles of $0^\circ$ and
$20^\circ$ are favored over larger values.  }
\label{fig:varybar}
\end{figure}

\section{Results and Discussions}

In our models, a bar develops self-consistently from the initially unbarred,
thin disk.  Bar formation enhances the radial streaming motions of disk
particles, so the radial velocity dispersion quickly grows much bigger than
the vertical one.  Consequently the disk buckles vertically out of the plane
like a fire hose; this is the well known buckling or corrugation instability
\citep{toomre_66,com_etal_90,rah_etal_91}.  It raises the vertical velocity
dispersion and increases the bar's thickness.  This happens on a short
dynamical timescale and saturates in a few hundred million years.  The central
part of the buckled bar is elevated well above the disk mid-plane and
resembles the peanut morphology of many bulges including the one in our Galaxy
\citep{com_etal_90,rah_etal_91}.

Out of a large set of $N$-body models, we find the one that best matches our
{\sl BRAVA} kinematic data after suitable mass scaling. The barred disk evolved from
a thin exponential disk that contains $M_{\rm d}=4.25\times 10^{10}\Ms$,
about 55 \% of the total mass at the truncation radius (5 scale-lengths).
The scale-length and scale-height of the initial disk are
$\sim$ 1.9~kpc and 0.2~kpc, respectively. The disk is rotationally supported and
has a Toomre-Q of 1.2.  The amplitude of the final bar is
intermediate between the weakest and strongest bars observed in galaxies. The
bar's minor-to-major axial ratio is about 0.5 to 0.6, and its half-length is
$\sim$ 4 kpc. Figure~1 (top three panels) shows face-on and side-on views of
the projected density of the best-fitting model. A distinctly peanut shaped
bulge is apparent in the edge-on projection.  Figure~1 (bottom
panel) shows the surface brightness distribution in Galactic coordinates as
seen from the Sun's vantage point.  Nearby disk stars dilute the peanut shape,
but the bar still looks boxy.  Moreover, from close up, an asymmetry in the
longitudinal direction is apparent; this means that the bar cannot be aligned
with the direction from the Sun to the Galactic center.  Rather, its near end
is at positive Galactic longitude, so it looks taller in that quadrant, and it
extends farther from the Galactic center on the near side than on the far
side.  Both the boxy shape and the asymmetry are in good agreement with the
morphology revealed by the COBE satellite near-infrared images
\citep{wei_etal_94,dwe_etal_95}.

Figure~2 compares the best-fitting model kinematics (solid
lines) with the mean velocity and velocity dispersion data from the
{\sl BRAVA} and other surveys \citep{ran_etal_09}.  All velocities presented
here have been converted to Galactocentric values (the line-of-sight
velocity that would be observed by a stationary observer at the Sun's
position).  For the first time, our model is able simultaneously to
match the mean velocities and velocity dispersions along two Galactic
latitudes ($-4^\circ$ and $-8^\circ$) and along the minor axis.

Figure~3 constrains the angle between the bar and the line that
connects the Sun to the Galactic center.  It compares the model
results with the data in the $b=-4^\circ$ major-axis strip as we vary
the above angle.  Clearly the smallest bar angles give the best match
to the velocity dispersions.  Intriguingly, we find that the velocity
dispersions provide much stronger constraints than the mean velocity
profile.  A bar angle of $0^\circ$ also matches the kinematics well.
However, the photometric asymmetry excludes a bar that is pointed at
the Sun.  We therefore conclude that the overall best-fitting model
has a bar angle of $\sim 20^\circ$.  Other studies
converged on a similar bar angle \citep{sta_etal_97,freude_98,fux_97,fux_99,bis_ger_02}.  The
excellent match to the data in Figures 1 -- 3 strongly supports the
suggestion that the boxy pseudobulge of the Milky Way is an edge-on,
buckled bar that evolved from a cold, massive disk. The thickened disk
in the pseudobulge-forming process may have contributed to
the thick disk of the Milk Way, as hinted from chemical similarities of Galactic bulge
and local thick disk stars \citep{alv_etal_10,ben_etal_10}.

The model in Figures 1 -- 3 contains no classical bulge component.  Could a
small classical bulge also be present?  Could it have been spun up by the
formation of a bar, flattened thereby and made hard to detect?  To constrain
such multi-component models with {\sl BRAVA} kinematics, we also constructed
models with a pre-existing classical bulge.  The distribution function for the
live classical bulge component was generated iteratively \citep{deb_sel_00} to
ensure that both the disk and the classical bulge were initially in
equilibrium, and we required that the bulge parameters are close to the
fundamental plane for classical bulges and ellipticals
\citep{kor_etal_09}. The setup of the disk is the same as in the disk-only
model.  We show in Figure~4 that inclusion of a classical bulge -- one widely
thought to be typical of Sbc spiral galaxies like our own -- greatly worsens
the model fit to the data.  The degradation is especially obvious along the
Galaxy's minor axis. Including a classical bulge with just 8\% of the disk
mass considerably worsens the fit of the model to the data. Our models rule
out that the Milky Way has a significant classical bulge whose mass is $>\sim$
15 \% of the disk mass.

\begin{figure*}[!ht]
\centerline{
\includegraphics[angle=0.,width=0.8\hsize]{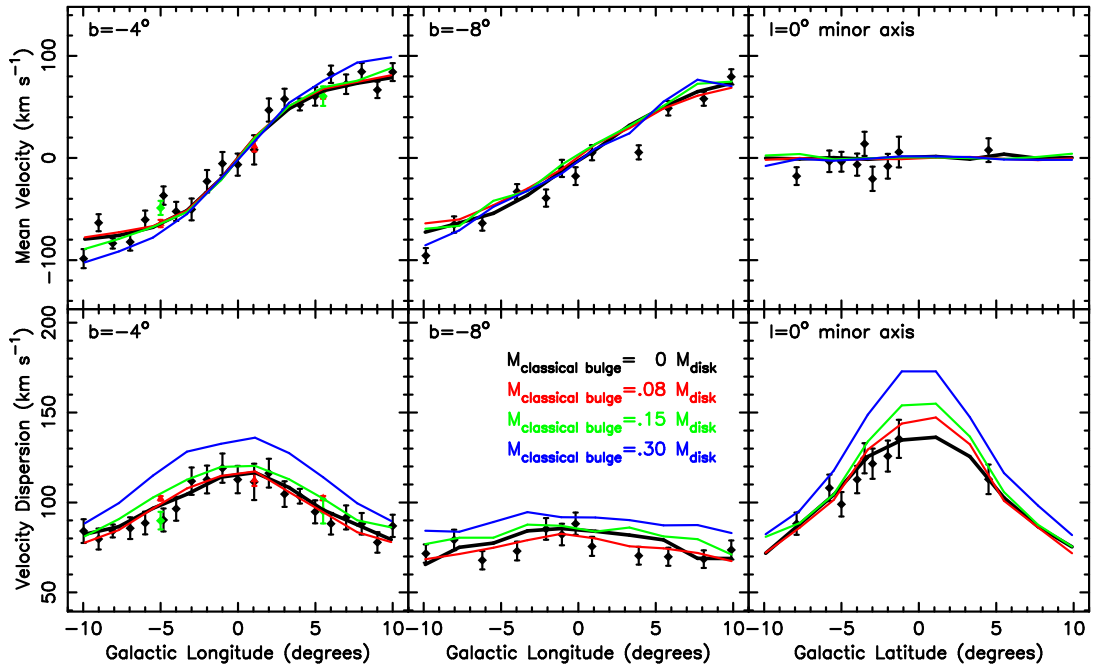}}
\caption{Fits to the kinematic data (cf.~Figure~2) of models that
include a pre-existing classical bulge. The heavy black lines from
Figure 2 represent the model without a classical bulge.  The red,
green, and blue lines are for models whose classical bulges have
masses of 8\%, 15\%, and 30\%, respectively, of the disk mass $M_{\rm
disk}$.  Including a classical bulge significantly worsens the model
fits to the data, especially along the minor axis. }
\label{fig:varybulge}
\end{figure*}

Could a smaller, merger-built bulge hide inside the boxy bar?  The only result
that we are aware of that might point to such a conclusion is the observed
drop in stellar metal abundances with increasing height above the Galactic
plane \citep{zoc_etal_08,zoccal_10}. \citet{zoc_etal_08} argue that this means
that the bulge must consist of both a classical and an edge-on bar component.
However, no kinematic gradient or transition corresponding to the abundance
gradient is observed.  Moreover, an abundance gradient can be produced within
the context of secular pseudobulge formation if some of the vertical
thickening is produced by resonant heating of stars that scatter off the bar
\citep{pfe_nor_90}.  If the most metal-poor stars are also the oldest stars,
then they have been scattered for the longest time and now reach the greatest
heights.

Our results have important implications for galaxy formation.  We demonstrate
that the boxy pseudobulge is not a separate component of the Galaxy but rather
is an edge-on bar.  Bars are parts of disks.  To be sure, the stars in our
Galactic bar are older than most disk stars. But those stars could have formed
over a short period of time but long before the bar structure formed
\citep{wyse_99,freema_08_IAU}, their old age
\citep{zoc_etal_03,ful_etal_07} is therefore not an argument
against the internal secular evolution model. Our kinematic observations show
no sign that the Galaxy contains a significant merger-made, ``classical''
bulge.  So, from a galaxy formation point of view, we live in a pure-disk
galaxy.  Our Galaxy is not unusual: it is very similar to another giant
edge-on galaxy with a boxy bulge, NGC~4565.  \citet{kor_bar_10} recently show
that NGC~4565 does not contain even a small classical bulge component and that
it therefore is another giant, pure-disk galaxy that contains no sign of a
merger remnant.  In fact, giant, pure-disk galaxies are common in environments
like our own that are far from rich clusters of galaxies \citep{kor_etal_10}.
Classical-bulge-less, pure-disk galaxies present an acute challenge to the
current picture of galaxy formation in a Universe dominated by cold dark
matter -- growing a giant galaxy via hierarchical clustering ($\Vc \simeq 220$
km~s$^{-1}$ in the Milky Way) involves so many mergers that it seems almost
impossible to avoid forming a substantial classical bulge
\citep{pee_nus_10,age_etal_10}.  How did our Galaxy grow so large with
no observational sign that it suffered a major merger after the time 9 -- 10
Gyr ago \citep[e.g.][]{win_kep_08} when the first disk stars formed?


\acknowledgements We thank Karl Gebhardt for
helpful discussions and warm support on this project.  We also acknowledge
support from the US National Science Foundation under grant AST-0607490 (JK)
and AST-0709479 (RMR).

\newcommand{\noopsort}[1]{} \newcommand{\singleletter}[1]{#1}

\end{document}